# Geo-neutrinos and Silicate Earth Enrichment of U and Th

S.T. Dye [a,b,*]
[a] Hawaii Pacific University, Kaneohe, Hawaii, USA
[b] University of Hawaii, Honolulu, Hawaii, USA

* Corresponding author.
*E-mail address:* sdye@phys.hawaii.edu (S.T. Dye).

ABSTRACT
The terrestrial distribution of U, Th, and K abundances governs the thermal evolution, traces the differentiation, and reflects the bulk composition of the earth. Comparing the bulk earth composition to chondritic meteorites estimates the net amounts of these radiogenic heat-producing elements available for partitioning to the crust, mantle, and core. Core formation enriches the abundances of refractory lithophile elements, including U and Th, in the silicate earth by ~1.5. Global removal of volatile elements potentially increases this enrichment to ~2.8. The K content of the silicate earth follows from the ratio of K to U. Variable enrichment produces a range of possible heat-producing element abundances in the silicate earth. A model assesses the essentially fixed amounts of U, Th, and K in the approximately closed crust reservoir. Subtracting these sequestered crustal amounts of U, Th, and K from the variable amounts in the silicate earth results in a range of possible mantle allocations, leaving global dynamics and thermal evolution poorly constrained. Terrestrial antineutrinos from β-emitting daughter nuclei in the U and Th decay series traverse the earth with negligible attenuation. The rate at which large subsurface instruments observe these geo-neutrinos depends on the distribution of U and Th relative to the detector. Geo-neutrino observations with sensitivity to U and Th in the mantle are able to estimate silicate earth enrichment, leading to a more complete understanding of the origin, accretion, differentiation, and thermal history of the planet.

## 1. Introduction

Detectable geo-neutrinos are electron antineutrinos from the decay series of terrestrial U and Th (Krauss et al., 1984). Large subsurface instruments efficiently detect geo-neutrino interactions in ultra-pure volumes of scintillating liquid. Presently, these detectors measure the rate and energy spectrum of geo-neutrino interactions but not geo-neutrino directions. Analyses compare detected interaction rates, as well as estimated numbers of interactions from U, and from Th, with predictions from geological models. The predicted rate of geo-neutrino interactions at a given detector location depends on the spatial distribution of U and Th abundances in the silicate earth. Two operating detectors observe a rate of geo-neutrino interactions consistent with predictions (Abe et al., 2008; Bellini et al., 2010). These observations currently lack the precision required to constrain geological models.

The major radiogenic heat-producing elements in the earth are U, Th, and K (Van Schmus, 1995). Modeling the bulk earth composition from measured chemical compositions of chondritic meteorites (Wasson and Kallemeyn, 1988) estimates the amounts of U, Th, and K available for partitioning to the crust, mantle, and core. Core formation and possible volatile element removal result in refractory lithophile element enrichment of the silicate earth between ~1.5 and 2.8 times chondritic abundances (McDonough and Sun, 1995; Palme and O'Neil, 2003; Lyubetskaya and Korenaga, 2007; Javoy et al., 2010). Variable enrichment suggests a range of U and Th abundances in the silicate earth, or primitive mantle. The K abundance follows that of U by an estimated K to U ratio. Enrichment of ~1.5 is representative of the earth accreting from enstatite chondrites (Javoy et al., 2010), which are relatively depleted of volatile elements. This predicts planetary radiogenic power of 11±1 TW. Greater enrichment entails removal of volatile elements from the primitive mantle, favoring a carbonaceous chondrite origin of earth. Models suggesting silicate earth enrichment of ~2.2 to 2.8 (McDonough and Sun, 1995; Palme and O'Neil, 2003; Lyubetskaya and Korenaga, 2007) predict planetary radiogenic power of 17±2 TW to 21±2 TW, respectively.

Geological investigations assess average U, Th, and K abundances stratified in the continental crust (Rudnick and Gao, 2003). Together with seismic determinations of the thickness and density of crustal layers (Bassin et al., 2000), model assessments of the essentially fixed amounts of these heat-producing elements result. This model estimates 8±2 TW of radiogenic power from U, Th, and K sequestered in the crust

Subtracting the crustal assessments from the enrichment-dependent estimates of U, Th, and K in the silicate earth predicts the depleted mantle allocations. These complimentary allocations establish the radiogenic contribution to the convective heat flux, influencing the dynamics and thermal evolution of the earth, and constraining secular cooling (Korenaga, 2008). Minimal silicate earth enrichment of 1.5 endows the depleted mantle with 3±2 TW of radiogenic power. The highest suggested enrichment of 2.8 raises the radiogenic power to 13±3 TW. Resolution of the abundances of U and Th in the depleted mantle estimates silicate earth enrichment, contributing information on the origin and thermal history of the earth.

The magnitude and distribution of U and Th abundances in the depleted mantle affects the rate of detectable geo-neutrinos. Calculation of the predicted rate assumes a radial symmetric mantle density distribution, as determined by seismology (Dziewonski and Anderson, 1981). The rate from a compositionally homogeneous mantle provides the reference. Concentrating a layer of heat-producing elements at the base of the mantle (Boyet and Carlson, 2005), called D", assumes a reduction of U and Th abundances in the overlying homogeneous mantle to values estimated for depleted mid-ocean ridge basalt mantle (Workman and Hart, 2005). This distribution lowers the geo-neutrino rate compared with the reference for a given silicate earth enrichment. Starting with enrichment-dependent primitive mantle U and Th abundances, subtracting the crustal assessments, and imposing the different mantle distributions on the remaining amounts, predicts a measurable range of geo-neutrino rates from the mantle. Measuring the mantle geo-neutrino rate with sufficient precision estimates silicate earth enrichment, constraining the accretion and differentiation of the planet.

Uncertainty in the measurement of the mantle geo-neutrino rate hinders resolution of the silicate earth enrichment and depends significantly on detector location. The main source of uncertainty is the crust model. Reducing this systematic uncertainty to allow measurement of silicate earth enrichment suggests several observational strategies. A much discussed strategy recommends observing mantle geo-neutrinos from the deep ocean basin far from the influence of the crust (Rothschild et al., 1998; Mantovani et al., 2004; Enomoto et al., 2007; Dye and Guillian, 2008). Alternative strategies using a continental observatory reduce systematic uncertainty by measuring the U and Th abundances in the local crust or by rejecting crustal geo-neutrinos by directional analysis.

This report begins with a brief review of neutrinos and the detection of geo-neutrinos, including a comparison of existing and proposed detectors. It synthesizes a spectrum of distributions of U and Th abundances in the depleted mantle by applying variable enrichment to a bulk earth of chondritic composition and assuming a crust model with fixed amounts of the heat producing elements. Next, it presents the predicted geo-neutrino signals and radiogenic powers for these synthesized distributions. This report continues with an exposition of geo-neutrino rate measurement errors associated with the existing and proposed detectors, including an investigation of lateral heterogeneities. Then it evaluates the two reported geo-neutrino observations. It explores possible geo-neutrino observations at selected detection sites, evaluating the sensitivity to the distribution of U and Th abundances in the depleted mantle. It concludes that geo-neutrino observations are capable of estimating silicate earth enrichment, advancing a more complete understanding of the origin, accretion, differentiation, and thermal history of the earth.

## 2. Geo-Neutrinos

Neutrinos are elementary particles, which interact dominantly via the weak force. This unique attribute follows from their apparent lack of electromagnetic properties (Giunti and Studenikin, 2009) and allows them to pass freely through the earth, virtually without attenuation. There are three observed types of neutrinos and their antiparticles, antineutrinos, each associated with one of the three charged leptons: electron, muon, and tau. Neutrinos from the earth and nuclear reactors correlate with the electron type.

Electron neutrinos ($\nu_e$) and antineutrinos ($\bar{\nu}_e$) emerge from unstable atomic nuclei during β decay, explaining the continuous spectrum of observed β-ray energies. Light proton-rich nuclei, which are produced in nuclear fusion reactions in the sun (Burbidge et al., 1957), undergo $β^+$ decay with the emission of an electron neutrino. Heavy neutron-rich nuclei, which are produced in a nuclear fission reactor or in a supernova (Burbidge et al., 1957), undergo $β^-$ decay with the emission of an electron antineutrino. At the surface of the earth, in the 1 – 10 MeV energy range, electron neutrinos from the sun dominate the neutrino flux, while electron antineutrinos from nuclear reactors and heavy element decay in the earth dominate the antineutrino flux. Although the solar neutrino flux is about an order of magnitude greater than the flux of terrestrial antineutrinos, these neutrinos and antineutrinos have different primary interactions with matter, allowing separate detection.

In the 1 – 10 MeV energy range, the detection of solar neutrinos utilizes elastic scattering on atomic electrons (Cravens et al., 2008), represented by

$$\nu_e + e^- \rightarrow \nu_e + e^-. \qquad (1)$$

The cross section for this process,

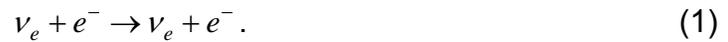

$$\sigma_{\nu_e e^- \rightarrow \nu_e e^-} \approx 9.5 E_\nu (MeV) \times 10^{-45} \text{ cm}^2, \qquad (2)$$

increases linearly with neutrino energy. Electrons scattered by this process point back to the sun, efficiently rejecting background in detectors capable of measuring the electron direction. The cross section for antineutrino-electron elastic scattering is smaller by a factor of 2.4.

A more prominent interaction for antineutrinos is inverse neutron decay, represented by

$$\bar{\nu}_e + p \rightarrow e^+ + n. \qquad (3)$$

The cross section for this process,

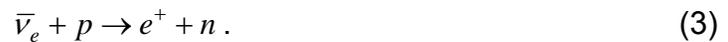

$$\sigma_{\bar{\nu}_e p \rightarrow e^+ n} \approx 9.3 [E_\nu (MeV)]^2 \times 10^{-44} \text{ cm}^2, \qquad (4)$$

increases with the square of the antineutrino energy, having a threshold energy of ~1.8 MeV. Detection of both interaction products efficiently rejects background.

The small interaction cross sections for weak interactions dictate large detection exposures. A unit of detector exposure for terrestrial antineutrino interactions is the $10^{32}$ proton-yr, corresponding to a detector with ~1.2 kT of instrumented hydrocarbon operated for 1 year. The inverse of this, measuring the interaction rate with 100% efficiency, is the terrestrial neutrino unit (TNU) (Mantovani et al., 2004).

Each neutrino type, electron, muon, and tau, is a quantum mechanical superposition of three distinct mass states: 1, 2, and 3. Mass state mixture determines the probability of observing a certain neutrino type. A neutrino of a given momentum has a periodically evolving mixture of mass states due to dispersion, causing the probability for observing a neutrino type to oscillate in passage (Gonzalez-Garcia and Nir, 2003). The probability for observing reactor or terrestrial antineutrinos is

$$P_{\bar{\nu}_e \to \bar{\nu}_e} \cong 1 - \sin^2(2\theta_{12})\sin^2(1.27\Delta m_{21}^2 L / E_{\bar{\nu}_e}), \qquad (5)$$

with $\theta_{12}$ and $\Delta m_{21}^2$ the measured oscillation parameters (Abe et al., 2008), $L$ the distance between source and detector in meters, and $E_{\bar{\nu}_e}$ the antineutrino energy in MeV. A 3 MeV antineutrino has an oscillation length of about 100 km, which is small compared with the spread in source distances for terrestrial antineutrinos. This justifies the use of the average survival probability 0.56 ± 0.02 for the suppression of the geo-neutrino flux at the detector (Enomoto et al., 2007). Calculation of the flux of reactor antineutrinos, which arrive at the detector from fixed source distances, requires use of the full expression for the survival probability.

The expected surface flux of antineutrinos in the 1 – 10 MeV range comes almost exclusively from the earth itself and commercial nuclear reactors. Contributions from other known sources, such as the diffuse supernova neutrino background and atmospheric neutrinos, are negligible in comparison. Variations of the surface flux are large, being strongly influenced by the local crust and proximity to nuclear reactors. Integrating the product of flux, interaction cross section, and neutrino oscillation survival probability over the energy spectrum gives the event rate (Mantovani et al., 2004; Enomoto et al., 2007; Fiorentini et al., 2007). Multiplying this rate by the detector exposure, given by the product of detector size and observation time, yields the expected number of antineutrino events observed with 100% efficiency. In practice, detectors observe background events mimicking inverse neutron decay. This study ignores these fake events, which primarily originate from secondary cosmic rays and ambient radioactivity.

Geo-neutrino detection typically uses the same technology employed by reactor antineutrino experiments for many decades (Reines and Cowan, 1953). In this technique, a surface array of inward-looking photomultiplier tubes monitors a large central volume of scintillating liquid. Antineutrinos interact dominantly with hydrogen nuclei, or free protons, in the scintillating liquid (Eq. 3). Correlated signals from the products of inverse neutron decay mark the interaction. Initially, a positron provides a measure of the geo-neutrino energy. The subsequent neutron capture, depositing fixed energy, tags the event.

The established antineutrino detection technique allows a spectral measurement of geo-neutrinos originating from the decay series of $^{238}$U, with energy 1.8 – 3.3 MeV, and $^{232}$Th, with energy 1.8 – 2.3 MeV. Terrestrial antineutrinos from all other radioactive isotopes, including $^{40}$K, lack the energy to initiate the inverse beta reaction on free protons. The highest energy geo-neutrinos derive only from $^{238}$U. This enables separate measurement of geo-neutrinos from the $^{238}$U and $^{232}$Th decay series, offering an estimate of the integrated thorium to uranium ratio (Raghavan et al., 1998; Rothschild et al., 1998). The traditional inverse beta coincidence technique provides limited information on geo-neutrino direction (Apollonio et al., 1999). Advances in detection technology offer improvement but not beyond the kinematic limits of the inverse neutron decay reaction, which favor the lower energy geo-neutrinos (Batygov, 2006). Measurement of geo-neutrino direction potentially allows discrimination of geo-neutrino source reservoirs and rejection of background from nuclear reactors.

Nuclear reactors produce antineutrinos with energies extending up to about 10 MeV (Djurcic et al., 2009), overlapping the detectable geo-neutrino spectrum at low energy. Their intensity and the oscillation parameters define the detection rate of reactor antineutrinos at a given site. Calculated detection rates result from 201 reactor complexes world-wide, operating with 90% duty cycle at a maximum collective power of 1.063 TW.

Large scintillating liquid instruments, typically deployed for physics and astrophysics investigations, detect geo-neutrino interactions. Their location and size critically determine the precision of geo-neutrino measurements. Two of the detectors, KamLAND at Kamioka, Japan (Abe et al., 2008), and Borexino at the Laboratorio Nazionale di Gran Sasso (LNGS) in L'Aquila, Italy (Bellini et al., 2010), are currently observing geo-neutrinos. A third detector, the former SNO refilled with scintillating liquid and called SNO+ at SNOLAB in Lively, Ontario, Canada (Chen, 2006), is preparing to resume operation. The remaining detectors in this study, Homestake at the Deep Underground Science and Engineering Laboratory (DUSEL) in Lead, South Dakota (Tolich et al., 2006), Baksan at the Baksan Neutrino Observatory (BNO) in the Russian Caucasus (Barabanov et al., 2009), LENA at the Center for Underground Particle Physics (CUPP) in Pyhasalmi, Finland (Hochmuth et al., 2006), and Hanohano in the Pacific Ocean (Dye et al., 2006), are proposed projects. Table 1 gives general information on the detector projects and the detection sites, including the calculated background rates from nuclear reactors.

## 3. Silicate Earth Enrichment

A scenario for the origin of the earth entails the accretion of chondritic meteorites. Chondrites, which are primitive undifferentiated meteorites, have compositional variations suggestive of assembly within the planetary nebula at different radii and epochs (Wasson and Kallemeyn, 1988). Comparing the bulk earth composition with a type of chondritic meteorite provides information on the origin and accretion of the planet. Enstatite chondrites and the earth appear to have comparable oxygen isotopic compositions as well as similarly small abundances of volatile elements and oxidized iron (Javoy, 1995). Within this group of meteorites, the high iron content of the EH class best accounts for the size of earth's core (Javoy, 1995). Alternatively, members of another group of chondritic meteorites, the carbonaceous chondrites, have different compositional similarities with the bulk earth, including enrichment of refractory elements, and depletion trends of moderately volatile elements (Palme and O'Neill, 2003). The chondritic meteorites with the highest volatile element abundances are the CI carbonaceous chondrites (Wasson and Kallemeyn, 1988). The CI and EH chondrite meteorites have similar U and Th abundances (Wasson and Kallemeyn, 1988). This study correlates these abundances with the primordial bulk earth composition, adopting values for U and Th of 8.2 (±10%) ppb and 29 (±5%) ppb, respectively, and setting the Th to U ratio equal to 3.5±0.4.

**Table 1**
Location, project status, latitude, longitude, and size of each large scintillating liquid detector are followed by the reactor background rate $r$.

|  | Location | Status | Lat. (°N) | Lon. (°E) | Free p (/$10^{32}$) | $r$ (TNU) |
|---|---|---|---|---|---|---|
| KamLAND | Kamioka | Operating | 36.43 | 137.31 | 0.6 | 208.2 |
| Borexino | LNGS | Operating | 42.45 | 13.57 | 0.1 | 31.4 |
| SNO+ | SNOLAB | Conversion | 46.47 | -81.20 | 0.6 | 46.6 |
| Homestake | DUSEL | Proposed | 44.35 | -103.75 | 0.5 | 9.4 |
| Baksan | BNO | Proposed | 43.29 | 42.70 | 4.0 | 11.3 |
| LENA | CUPP | Proposed | 63.66 | 26.05 | 36.7 | 22.6 |
| Hanohano | Pacific | Proposed | 19.72 | -156.32 | 7.3 | 1.3 |

Chemical differentiation of the primordial earth proceeds with the formation of a metallic core, primarily Fe and Ni, enveloped by a silicate shell or primitive mantle. Assuming negligible amounts of U and Th enter the core, this process increases the chondritic abundances in the primitive mantle by a factor of 1.48. Upper bounds on the concentration of U in the core from laboratory experiments reduce U enrichment either insignificantly (Wheeler et al., 2006) or to 1.44 (Malavergne et al., 2007). Comparing Th/U between the mantle and chondrites suggests core concentrations lowering U enrichment to 1.19 (Labrosse et al., 2001). A similar constraint obtains from allowing ~1 TW of radiogenic power due to U in the core (Anderson, 2002). Although suggestions for U in the core could push silicate earth enrichment of U as low as ~1.2, Th appears to be virtually unaffected (Labrosse et al., 2001). This study assumes negligible amounts of U and Th reside in the core (McDonough, 2003), imposing a lower bound on silicate earth enrichment of ~1.5.

Enrichment of U and Th beyond that of core formation could result from removal of volatile elements from the primitive mantle. This would not involve a primordial earth built from enstatite chondrites, as these meteorites are deficient in volatile elements (Wasson and Kallemeyn, 1988). Because carbonaceous chondrites have the highest volatile element abundances, these meteorites provide a convenient reference for silicate earth enrichment. Models of the primitive mantle composition offer a range of enrichment relative to CI meteorites, starting as low as ~2.2±0.4 (Lyubetskaya and Korenaga, 2007) and extending up to ~2.8 (McDonough and Sun, 1995; Palme and O'Neil, 2003).

Radiogenic power, particularly in the mantle, has a significant impact on the dynamics and thermal evolution of the earth (Jaupart et al., 2007; Korenaga, 2008). Estimates of the radiogenic power in a given geochemical reservoir follow from the mass of the reservoir and the abundances and specific isotopic heat productions of U, Th, and K (Van Schmus, 1995). Table 2 compares the abundances of heat-producing elements and the Th to U ratios in chondrites with those predicted by primitive mantle models and presents estimates of the silicate earth radiogenic power for each model. An assessment of the U and Th abundances in the primitive mantle would gauge silicate earth enrichment and radiogenic power, constraining the accretion and secular cooling of the earth.

**Table 2**
Approximate uranium, thorium and potassium-40 abundances of chondrites and the primitive mantle according to bulk silicate earth models are compared along with the radiogenic powers. Bulk silicate earth mass is $4.03 \times 10^{27}$ g.

|  | U (ng/g) | Th (ng/g) | Th/U | K/U | $^{40}$K (ng/g) | P (TW) |
|---|---|---|---|---|---|---|
| Chondrites | 8±1 | 29±1 | 3.5±0.4 | ------ | 67-95 | ------ |
| Javoy et al., 2010 | 12±1 | 44±2 | 3.5±0.4 | 12,000 | 18±4 | 11±1 |
| Lyubetskaya & Korenaga, 2007 | 17±3 | 63±11 | 3.6±0.9 | 11,000 | 23±5 | 16±3 |
| McDonough & Sun, 1995 | 20±4 | 80±12 | 3.9±1.0 | 11,800 | 33±7 | 20±4 |
| Palme & O'Neill, 2003 | 22±3 | 83±13 | 3.8±0.8 | 11,900 | 31±5 | 21±3 |

## 4. Crust and Mantle Models

Chemical differentiation of the earth continues with extraction of the continental crust from the primitive mantle. The continental crust is a major reservoir of U, Th, and K, containing at least one-third of the planetary budget of these heat-producing elements (Jaupart et al., 2007). Although the radiogenic power of the continental crust contributes to the planetary heat flux, it does not participate in the dynamically important convective heat flux of the mantle (Korenaga, 2008). Indeed, continents act like thermal blankets, mitigating mantle heat loss (Grigne and Labrosse, 2001). A model estimates the amounts of U, Th, and K in the crust to arrive at the depleted mantle allocations of U, Th, and K, which are constrained by silicate earth enrichment.

The crust model comprises a comprehensive assessment of the average composition of the continental crust (Rudnick and Gao, 2003) and seismic measurements of the thicknesses, densities, and lateral extent of crustal layers (Bassin et al., 2000). This model estimates the masses of U, Th, and K in the crust, predicting an average Th to U ratio of 4.3±1.2 and producing radiogenic power of 7.8±1.5 TW. Table 3 specifies the model values for the abundances of the heat-producing elements, Th to U ratio, and radiogenic power for the oceanic crust, sediments, and upper, middle, and lower layers of continental crust.

**Table 3**
Model information for the crust is specified according to type and stratified layers.

| Reservoir | M ($10^{25}$ g) | U (µg/g) | Th (µg/g) | Th/U | $^{40}$K (µg/g) | P (TW) |
|---|---|---|---|---|---|---|
| Bulk crust | 2.79 | 1.17±.30 | 4.97±.66 | 4.3±1.2 | 1.55±.26 | 7.8±1.5 |
| CC upper | 0.69 | 2.70±.57 | 10.5±1.1 | 3.9±0.9 | 2.80±.23 | 4.22±0.61 |
| CC middle | 0.71 | 1.30±.40 | 6.50±.52 | 5.0±1.6 | 2.30±.32 | 2.55±0.43 |
| CC lower | 0.65 | .20±.16 | 1.20±.96 | 6.0±6.8 | .60±.48 | 0.44±0.35 |
| CC sediments | 0.08 | 2.70±.57 | 10.5±1.1 | 3.9±0.9 | 2.80±.23 | 0.50±0.07 |
| OC | 0.63 | .10±.02 | .22±.02 | 2.2±0.5 | .15±.02 | 0.12±0.02 |
| OC sediments | 0.03 | 1.70±.34 | 6.90±.69 | 4.1±0.9 | 1.70±.17 | 0.12±0.02 |

The abundances of U, Th, and K in the depleted mantle and their distribution control convection in this reservoir, influencing the evolution of the planet (Stevenson, 2003). A range of abundances result from subtracting the relatively well known contents sequestered in the crust from the enrichment-dependent amounts in the primitive mantle. Primitive mantle modeling assumes K to U of 12,000 (±15%). Variable silicate earth enrichment affects the radiogenic power of both the primitive mantle and the depleted mantle. Adjusting enrichment from 1.5 to 2.8 increases the radiogenic power of the primitive mantle from 11±1 TW to 21±2 TW and that of the depleted mantle from 3±2 TW to 13±3 TW, respectively. Dividing the enrichment-dependent power of the primitive and depleted mantle by the planetary heat flux of 46±3 TW (Jaupart et al., 2007), calculates the bulk earth Urey ratio and the dynamically important convective Urey ratio, respectively (Korenaga, 2008). Figure 1, which presents these results, demonstrates the effect of silicate earth enrichment on radiogenic power.

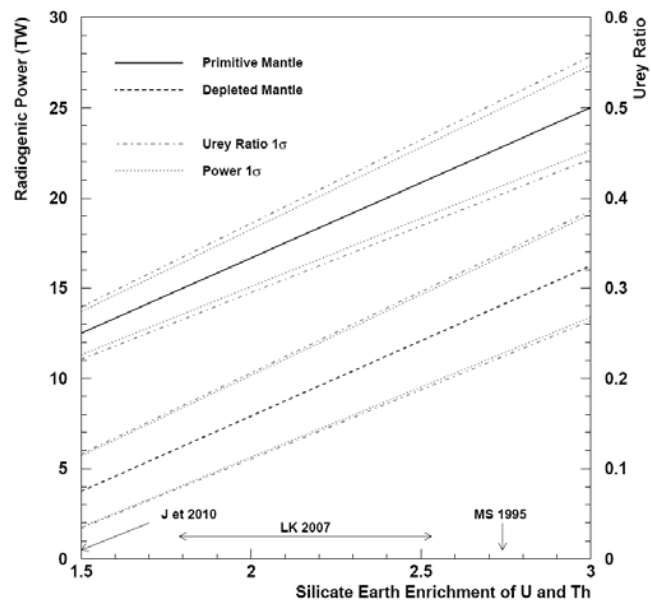

**Fig. 1.** Radiogenic power and Urey ratio of the primitive and depleted mantle are plotted as a function of silicate earth enrichment. Model predictions for silicate earth enrichment are shown for comparison.

Mantle mixing depends on the distribution of heat-producing elements in this geochemical reservoir. Tomographic images of subducting slabs suggest deep mantle convection (van der Hilst et al., 1997), disfavoring models with upper and lower mantle reservoirs separated at the 670 km seismic discontinuity (Kramers and Tolstikhin, 1997; Turcotte et al., 2001). Measured differences in Nd isotopic ratios between chondrites and terrestrial rocks suggest the presence of an enriched layer isolated in the D" region just above the core (Boyet and Carlson, 2005). This evidence favors models of whole mantle convection driven by radioactive elements concentrated at the base of the mantle (Tolstikhin and Hofmann, 2005; Tolstikhin et al., 2006; Boyet and Carlson, 2006). Although the distribution of the net amounts of U, Th, and K in the depleted mantle does not affect the convective Urey ratio, it does influence the geo-neutrino flux at the surface of the earth.

The amounts of heat-producing elements available for concentration in the D" region increases with silicate earth enrichment. Assessment of these amounts requires assignment of the minimum allowable abundances of U, Th, and K in the depleted mantle. Several estimates of the depleted mantle composition are available for this purpose (Salters and Stracke, 2004; Workman and Hart, 2005). To fully explore the maximum effects of a concentrated D" region, this study adopts the estimate with the lowest abundances of heat-producing elements, which specifies a depleted mantle Th to U ratio of ~2.5 (Workman and Hart, 2005).

## 5. Geo-neutrino rates from the core, mantle, and crust

Calculation of the expected rate of observable geo-neutrino interactions follows directly from the model distribution of U and Th in the core, mantle, and crust relative to the detector location (Mantovani et al., 2004; Enomoto et al., 2007). Observation of geo-neutrinos from the core seems a remote possibility. An upper bound on the concentration of U in the core from laboratory experiments limits the rate of geo-neutrinos to <0.3 TNU (Malavergne et al., 2007). Comparing Th/U between the mantle and chondrites extends the limit to <2 TNU (Labrosse et al., 2001). A similar constraint obtains from allowing 1 TW of radiogenic power due to U in the core (Anderson, 2002). Assuming negligible amounts of U and Th in the core (McDonough, 2003) restricts detectable geo-neutrino production to the mantle and crust.

The predicted rate of observable geo-neutrino interactions near the surface of the earth depends strongly on proximity to continental crust. According to silicate earth models, this enriched reservoir of U and Th is the dominant source of geo-neutrinos at continental locations. Moving offshore to the deep ocean basin shifts the dominance to the mantle. Table 4 details the rates of observable geo-neutrino interactions predicted by the crust model at the detector locations specified in Table 1. The model estimates a crust rate of ~40 TNU with an uncertainty of 24% on continents, decreasing to ~3 TNU with an uncertainty of 18% in the deep ocean basin.

**Table 4**
The rates in TNU of geo-neutrino events from the bulk crust and each sub-reservoir are presented for the detection sites. Uncertainties derive from the crustal model. Uncertainty in the U and Th abundances in the upper continental crust is the dominant source of uncertainty in the bulk crust rate prediction. The crustal Th/U ratio measured by geo-neutrinos is 4.2 (+0.9/-0.5) for all sites except Pacific, where it is 3.9 (+0.3/-0.5).

| Reservoir | Kamioka | LNGS | SNOLAB | DUSEL | BNO | CUPP | Pacific |
|---|---|---|---|---|---|---|---|
| Bulk Crust | 23.9±5.7 | 30.0±7.2 | 36.2±8.7 | 41.1±9.5 | 41.2±9.6 | 38.1±9.3 | 3.3±0.6 |
| CC upper | 13.8±2.7 | 16.6±3.3 | 22.0±4.3 | 23.7±4.7 | 23.0±4.5 | 22.6±4.5 | 1.5±0.3 |
| CC middle | 7.0±1.8 | 8.8±2.2 | 11.4±2.9 | 12.4±3.1 | 11.7±2.9 | 12.3±3.1 | 0.8±0.2 |
| CC lower | 1.2±1.0 | 1.4±1.1 | 1.7±1.4 | 1.4±1.1 | 1.4±1.2 | 1.9±1.6 | 0.1±0.1 |
| Sediments | 1.5±0.3 | 2.9±0.6 | 0.9±0.2 | 3.5±0.7 | 5.1±1.0 | 1.1±0.2 | 0.2±0.0 |
| OC | 0.3±0.0 | 0.3±0.0 | 0.2±0.0 | 0.2±0.0 | 0.1±0.0 | 0.2±0.0 | 0.7±0.1 |

Assuming a radial symmetric mantle, the rate of observable geo-neutrino interactions from this depleted reservoir is uniform near the surface of the planet. The predicted rate depends on the enrichment of the silicate earth and the distribution of U and Th within the depleted mantle. For a homogeneous depleted mantle without a concentrated D" region, a silicate earth enrichment of 1.5 produces a mantle geo-neutrino rate of 3±2 TNU, increasing to 12±2 TNU for an enrichment of 2.8. Because the crust has a higher Th to U ratio than chondrites, the depleted mantle has a lower Th to U ratio than chondrites. As the silicate earth enrichment rises from 1.5 to 2.8, the Th to U ratio in the depleted mantle increases from ~2.1 to 3.1. Because the depleted mantle is compositionally homogeneous in this model, the scaled geo-neutrino Th to U ratio is equivalent.

Sequestering a portion of the heat-producing elements in the D" region reduces the expected mantle geo-neutrino rate by moving U and Th farther from the surface. A silicate earth enrichment of 1.5, affording little excess U and Th for the D" region, has no significant effect on the mantle geo-neutrino rate. However, an enrichment of 2.8, allowing a concentrated D" region, reduces the expected mantle geo-neutrino rate by ~2 TNU. Concentrating refractory lithophile elements in the D" region affects insignificantly the geo-neutrino measurement of the Th to U ratio, causing at most a slight downward trend. Figure 2 displays the mantle geo-neutrino Th to U ratio and rate as a function of silicate earth enrichment. Significant differences in the modeled geo-neutrino quantities reveal considerable sensitivity to silicate earth enrichment. Experimental resolution of silicate earth enrichment, however, depends on the systematic and statistical uncertainties of the geo-neutrino measurements.

### 6. Geo-neutrino measurement uncertainties

The precision of the geo-neutrino measurements at a given observation site determines the capability for resolving silicate earth enrichment. Both statistical and systematic uncertainties affect the precision. Statistical errors depend on the total antineutrino rate and exposure. Systematic errors depend on instrumental uncertainties, including detector exposure and antineutrino energy measurement, on model uncertainties in the distribution of terrestrial U and Th abundances, and on the antineutrino rate from nuclear reactors. Measurements of geological interest include the total geo-neutrino rate ($g$), crust rate ($c$), mantle rate ($m$), and the ratio Th to U rates, appropriately scaled to give the ratio of Th to U abundances.

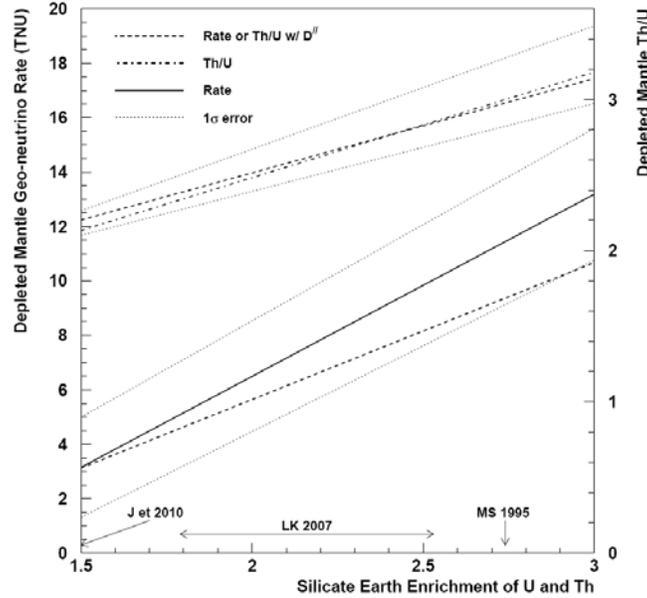

**Fig. 2.** Geo-neutrino rate and Th to U ratio from the depleted mantle are plotted as a function of silicate earth enrichment. Model predictions for silicate earth enrichment are shown for comparison.

The precision of the geo-neutrino measurements at a given observation site determines the capability for resolving silicate earth enrichment. Both statistical and systematic uncertainties affect the precision. Statistical errors depend on the total antineutrino rate and exposure. Systematic errors depend on instrumental uncertainties, including detector exposure and antineutrino energy measurement, on model uncertainties in the distribution of terrestrial U and Th abundances, and on the antineutrino rate from nuclear reactors. Measurements of geological interest include the total geo-neutrino rate ($g$), crust rate ($c$), mantle rate ($m$), and the ratio Th to U rates, appropriately scaled to give the ratio of Th to U abundances.

The rate of antineutrino events in the detectable geo-neutrino energy range ($n$) is simply the sum of rates from the crust, mantle, and nuclear reactors ($r$)

$$n = c + m + r. \qquad (6)$$

In the following equations for the uncertainties of the total and mantle rates of geo-neutrinos, the first term on the right side of the equations accounts for the statistical error and the remaining terms contribute to the systematic error.

$$\delta g = [n/\varepsilon + (n/\varepsilon)^2 (\delta \varepsilon)^2 + (\delta r)^2]^{1/2} \qquad (7)$$

$$\delta m = [n/\varepsilon + (n/\varepsilon)^2 (\delta \varepsilon)^2 + (\delta r)^2 + (\delta c)^2]^{1/2} \qquad (8)$$

Whereas experimental values establish the uncertainties in the exposure and the nuclear reactor rate at ~2% (Abe at al., 2008), the crust model prescribes a much larger uncertainty in the crust rate of ~25%. Without a significant reduction in this systematic uncertainty in the crust rate, continental geo-neutrino observatories are unable to resolve silicate earth enrichment.

Measurement of antineutrino energy enables separate estimates of geo-neutrinos from the $^{238}$U and $^{232}$Th decay series, providing information on the terrestrial Th to U ratio. These estimates require splitting the event rate ($n$) in the geo-neutrino energy range (1.8 – 3.3 MeV) at the endpoint energy of the Th decay series (2.3 MeV). At higher energy (2.3 – 3.3 MeV) the rate is only due to U geo-neutrinos and reactor antineutrinos ($n_H = g_H + r_H$), while at lower energy (1.8 – 2.3 MeV) the rate is due to U and Th geo-neutrinos as well as reactor antineutrinos ($n_L = g_L + r_L$). Operationally, this splitting necessitates calculation of the fractions of the U geo-neutrino ($f_U = 0.54 \pm 0.01$) and the reactor antineutrino ($f_r = 0.86 \pm 0.03$) event rates with energy greater than the Th decay series endpoint. Uncertainty in these fractions depends on knowledge of the energy scale and oscillation parameters (Abe et al., 2008). The definitions above permit the expression of the rate of all geo-neutrino events ($g = n_H + n_L - r$) and the rate of geo-neutrino events at higher energy ($g_H = n_H - r f_r$) in terms of measured and calculated quantities. Differences in the U and Th activities and the decay series β energies cause the ratio of the Th to U geo-neutrino rates to differ from the ratio to Th to U abundances by a calculable factor ($\alpha = 15.2$). Using these definitions, the expression for the ratio of Th to U abundances is

$$Th/U = \alpha \left\{ f_U \left( \frac{g}{g_H} \right) - 1 \right\}. \qquad (9)$$

The statistical and systematic uncertainties of the Th to U ratio are separately

$$\delta(Th/U)_{stat} = \frac{\alpha f_U}{g_H^2} \left( \frac{n_H g_L^2 + n_L g_H^2}{\varepsilon} \right)^{1/2} \qquad (10)$$

$$\delta(Th/U)_{sys} = \frac{\alpha f_U}{g_H^2} \left\{ \left[ g g_H \left( \frac{\delta f_U}{f_U} \right) \right]^2 + \left[ g r_H \left( \frac{\delta f_r}{f_r} \right) \right]^2 + (n r_H - r n_H)^2 \left[ \left( \frac{\delta \varepsilon}{\varepsilon} \right)^2 + \left( \frac{\delta r}{r} \right)^2 \right] \right\}^{1/2}. \qquad (11)$$

Uncertainty in the fractions of the event rates with energy above the Th decay series endpoint dominates the systematic error. At observation sites where the reactor rate is comparable to the geo-neutrino rate, uncertainty in the reactor rate fraction limits the precision to ~±1.3. At sites where the reactor rate is several times smaller than the geo-neutrino rate, uncertainty in the geo-neutrino rate fraction limits the precision to about ~±0.5.

Table 5 summarizes the systematic uncertainties for each detection site calculated with an assumed mantle rate of 12 TNU. Uncertainty in the crust model is the dominant source of systematic error for geo-neutrino rate measurements. At continental sites, this uncertainty is comparable in magnitude to the highest expected mantle rate. Lacking more precise estimates of the abundances of U and Th distributed in the crust, particularly in the region local to the detector, resolving silicate earth enrichment with continental geo-neutrino observations is not possible. The situation is significantly different with oceanic observations. At the Pacific site, the systematic uncertainty in the measured mantle geo-neutrino rate falls to about 25% of the lowest expected rate. A projected estimate of the silicate enrichment from the Pacific site carries a similar precision.

**Table 5**
Systematic errors in TNU of event rates from the bulk crust, nuclear reactors, and detection exposure are listed for each observation site. These errors are used to calculate the systematic errors of the total geo-neutrino rate and the mantle rate, according to Eq. 7 and 8, respectively. The final entry is the estimated systematic uncertainty in the ratio of Th to U abundances. In calculations of the systematic uncertainty in detection exposure and ratio of Th to U abundances, a mantle rate of 12 TNU is used, corresponding to silicate earth enrichment of 2.8.

| Source | Kamioka | LNGS | SNOLAB | DUSEL | BNO | CUPP | Pacific |
|---|---|---|---|---|---|---|---|
| Bulk crust, $\delta c$ | 5.7 | 7.2 | 8.7 | 9.5 | 9.6 | 9.3 | 0.6 |
| Nuclear reactor, $\delta r$ | 4.2 | 0.6 | 0.9 | 0.2 | 0.2 | 0.5 | 0.0 |
| Exposure, $n\delta\varepsilon/\varepsilon$ | 4.9 | 1.5 | 1.9 | 1.3 | 1.3 | 1.5 | 0.3 |
| Geo-neutrino, $\delta g$ | 6.4 | 1.6 | 2.1 | 1.3 | 1.3 | 1.5 | 0.3 |
| Mantle, $\delta m$ | 8.6 | 7.4 | 9.0 | 9.6 | 9.7 | 9.4 | 0.7 |
| $\delta$(Th/U) | 9.3 | 1.3 | 1.6 | 0.5 | 0.5 | 0.8 | 0.4 |

A potentially significant source of systematic uncertainty not included in the results presented in Table 5 is lateral heterogeneity of the terrestrial U and Th distributions. Whereas oceanic crust is quite uniform in composition (Taylor and McLennan, 1985), lateral heterogeneity certainly exists in the continental crust, as evidenced by uranium mining at specific locations. It exists in the mantle at least at the scale of subducting slabs (van der Hilst et al., 1997). Larger scale mantle heterogeneity is possibly associated with anomalously shallow regions of the seafloor (McNutt and Judge, 1990).

An investigation of mantle heterogeneity calculates the increase in the geo-neutrino rate due to a spherical volume with U and Th abundances twice those of an otherwise homogeneous mantle. The spherical volume has the same seismically-determined density profile as the mantle (Dziewonski and Anderson,

1981). Two sizes of spheres are under consideration. Each has a mass <2% that of the entire mantle, insignificantly perturbing the total amounts of U and Th. A large sphere, tangent to the core-mantle boundary and the base of the lithosphere, increases the mantle geo-neutrino rate at the surface directly above the sphere by ~38%. A smaller sphere, also tangent to the core-mantle boundary but extending to the 670-km seismic transition zone, increases the mantle geo-neutrino rate at the surface directly above the sphere by ~9%. Moving off the axis running through the top of the sphere and the center of the earth diminishes the effect by roughly a factor of two every 10°. Mantle heterogeneity of the magnitude injected by the large sphere has a significant effect on the near-axis mantle geo-neutrino rate, potentially skewing the estimate of silicate earth enrichment. A candidate location for such lateral heterogeneity of U and Th abundances in the mantle is the South Pacific superswell (McNutt and Judge, 1990). A comparison of geo-neutrino measurements above this superswell and the abyssal plain in the Northeast Pacific would assess mantle heterogeneity.

By using average U and Th concentrations, the crust model fails to account for lateral heterogeneity. Only the crust near the detector has a significant effect on the geo-neutrino rate (Mantovani et al., 2004; Enomoto et al., 2007). According to the crust model, roughly 90% of the crustal geo-neutrino rate originates within 2000 km of the continental detector. About two thirds of this rate derives from the upper crust and sediments. Refining the crust model through comprehensive study of the U and Th distributions local to the detector reduces the influence of lateral heterogeneity (Enomoto et al., 2007). A complimentary investigation assesses the U and Th content of a crustal column centered on a borehole with heat flow measurements (Perry et al., 2009).

## 7. Reported geo-neutrino observations

Two operating detectors observe a total rate of geo-neutrino events consistent with the crust model and all values of the silicate earth enrichment between 1.5 and 2.8 (Abe et al., 2008; Bellini et al., 2010). Operators of the longest running project, which provided the first observation of geo-neutrinos (Araki et al., 2005), report a rate of 29±11 TNU from a $2.44 \times 10^{32}$ proton-yr exposure of a detector in Japan (Abe et al., 2008). The crust model with silicate earth enrichments of 1.5 and 2.8 predict total geo-neutrino rates of 27±6 TNU and 36±6 TNU, respectively. A more recent project reports a rate of 65 (+27/-22) TNU from a $0.15 \times 10^{32}$ proton-yr exposure of a detector in Italy (Bellini et al., 2010). The crust model with silicate earth enrichments of 1.5 and 2.8 predict total geo-neutrino rates of 33±7 TNU and 42±8 TNU, respectively. Without reducing systematic uncertainty in the predicted crust rate, it is unlikely that further operation of either of these detector projects can constrain silicate earth enrichment. Additionally, nuclear reactors interfere with the measurement in Japan, while slowly accruing exposure hampers the estimate in Italy. Nonetheless, these pioneering observations of geo-neutrinos demonstrate the viability of the detection technique and guide the development of projects capable of estimating silicate earth enrichment.

## 8. Resolving silicate earth enrichment

Variable enrichment of a chondritic earth allocates significantly different amounts of U and Th to the depleted mantle. Both the rate of geo-neutrinos from the mantle and the ratio of Th to U geo-neutrino rates from the mantle grow with increasing silicate earth enrichment (Fig. 2). Measuring these quantities with sufficient precision estimates silicate earth enrichment, constraining models of the origin and thermal evolution of the earth. In principle, these measurements are possible at any site with adequate control of background. Background levels in the deep ocean basin, several thousand kilometers from continental crust and nuclear reactors, are a small fraction of the lowest predicted mantle signal. Achieving similar sensitivity at a continental site requires extra measures to reduce background.

The systematic uncertainty in the background-subtracted mantle rate for the Pacific geo-neutrino observatory is ±0.7 TNU, dominated by the crust model (Table 5). Producing a statistical uncertainty of the same magnitude requires at most a 34 x$10^{32}$ proton-yr exposure. The full uncertainty of ~±1 TNU corresponds to a precision of ±0.3 in silicate earth enrichment at the 95% confidence level for a homogeneous mantle, degrading slightly to ±0.4 (95% CL) for a concentrated layer at the base of the mantle. Lateral heterogeneity of U and Th abundances in the mantle introduces a potentially large source of systematic uncertainty in the mantle geo-neutrino rate. However, deploying a movable detector (Learned et al., 2008) at multiple deep ocean locations, such as above the South Pacific superswell and the Northeast Pacific abyssal plain, assesses lateral heterogeneity. Systematic uncertainty in the homogeneous mantle geo-neutrino Th to U ratio at the Pacific site is ±0.4, dominated by error in the detector energy scale (Table 5). The exposure yielding an equivalent statistical uncertainty is 200x$10^{32}$ proton-yrs. The full uncertainty of ~±0.6 spans the entire range of variation in the Th to U ratio. A measurement of the mantle Th to U ratio capable of constraining silicate earth enrichment requires better knowledge of the detector energy scale and demands a large exposure.

At a continental site remote from nuclear reactors, the dominant source of background to mantle geo-neutrinos is the continental crust. Possible methods for reducing this background include measuring the U and Th abundances in the crust local to the detector and measuring geo-neutrino direction. The continental crust near the detector dominates the geo-neutrino rate. Reducing the model uncertainties in the U and Th abundances in the local crust has the largest impact. For example, the model predicts ~80% of the 40 TNU crustal geo-neutrino rate originating within a radius of ~1000 km. The remaining 20% of the crustal geo-neutrino rate carries the ~±25% model uncertainty, contributing an error of ~±5% to the total crustal rate. Reducing the uncertainty in the geo-neutrino rate from the local crust to ~±6% through geological measurements contributes another ~±5% to the total crust rate error. A 22x10$^{32}$ proton-yr exposure would introduce statistical uncertainty of ~±1 TNU. The resulting total uncertainty of ±3 TNU translates to a precision of ±0.9 (95% CL) in the silicate earth enrichment. A corresponding measurement of the Th to U ratio, sampling crust and mantle roughly equally, carries an uncertainty of ~±0.8, dominated by reactor antineutrinos. This uncertainty potentially spans the entire range of values, 2.1 to 3.2, predicted by variable silicate earth enrichment.

Measuring geo-neutrino direction, requiring advances in detection technology, provides another method for reducing background from continental crust. At a continental site, most geo-neutrinos from the crust travel in a near horizontal direction, while those from the mantle travel in an upward-going direction. Assume, for example, an angular resolution characterized by a normal distribution with 68% probability of measuring geo-neutrino direction to within ±30°. Selecting measured directions greater than 45° below the horizon then rejects ~92% of the crust geo-neutrinos and ~50% of the mantle geo-neutrinos. Figure 3 displays simulated distributions of directions at a continental site for a silicate earth enrichment of 1.5. Using the assumed angular resolution, the proposed 45° cut and a 20x10$^{32}$ proton-yr exposure, results in an uncertainty of ~±3 TNU. This translates to a precision of ~±0.8 (95% CL) in the silicate earth enrichment.

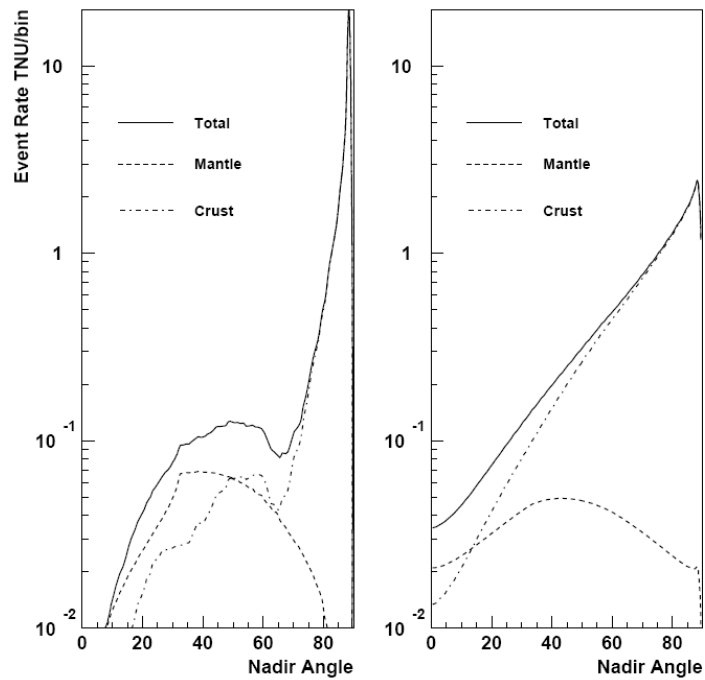

**Fig. 3.** Predicted geo-neutrino event rates at a continental site are plotted as a function of the nadir angle, which is measured with respect to the direction straight down. The right panel shows the direction distributions as calculated. The right panel shows the direction distributions as measured with angular resolution of ±30°.

## 9. Conclusions

Modeling increasing enrichments of refractory lithophile elements in a chondritic earth amplifies abundances of U and Th in the depleted mantle. Geo-neutrino observations are capable of measuring this silicate earth enrichment, addressing the origin, dynamics and thermal evolution of the planet. A silicate earth enrichment of 1.5, suggesting an enstatite chondrite earth, predicts a convective Urey ratio of 0.07±0.04 and a mantle geo-neutrino rate of 3±2 TNU. An enrichment of 2.8, heavily favoring a carbonaceous chondrite earth, increases the predicted ratio to 0.29±0.06 and rate to 12±2 TNU. Observations of mantle geo-neutrinos from the Pacific Ocean can estimate silicate earth enrichment with a precision of ~±0.3 (95% CL), discriminating between geological models. Observations from continents require significant reduction in the systematic uncertainty in the geo-neutrino rate from the crust to potentially estimate enrichment with a precision of ~±0.8 (95% CL). Although the mantle geo-neutrino Th to U ratio is sensitive to silicate earth enrichment, measurements require potentially excessive exposures and resulting uncertainties are probably too large to discriminate between models.

## Acknowledgements
The author thanks Rick Carlson, John Mahoney, Bill McDonough, Roberta Rudnick, and Bill White for comments and/or discussions. This work was supported in part by a grant (#0855838) from the National Science Foundation.